\documentclass{DISproc}
\usepackage[utf8x]{inputenc}
\usepackage[english]{babel}
\usepackage{subfigure}
\usepackage{mathrsfs}
\if@mathematic
   \def\vec#1{\ensuremath{\mathchoice
                     {\mbox{\boldmath$\displaystyle\mathbf{#1}$}}
                     {\mbox{\boldmath$\textstyle\mathbf{#1}$}}
                     {\mbox{\boldmath$\scriptstyle\mathbf{#1}$}}
                     {\mbox{\boldmath$\scriptscriptstyle\mathbf{#1}$}}}}
\else
   \def\vec#1{\ensuremath{\mathchoice
                     {\mbox{\boldmath$\displaystyle#1$}}
                     {\mbox{\boldmath$\textstyle#1$}}
                     {\mbox{\boldmath$\scriptstyle#1$}}
                     {\mbox{\boldmath$\scriptscriptstyle#1$}}}}

\begin{document}
\title{Two-Loop Gluon Regge Trajectory from\\ Lipatov's Effective Action}

\author{{\slshape Grigorios Chachamis$^1$, Martin Hentschinski$^2$, José Daniel Madrigal$^2$,  Agustín Sabio Vera$^2$}\\[1ex]
$^1$Paul Scherrer Institut, CH-5232 Villigen PSI, Switzerland\\
$^2$Instituto de Física Teórica UAM/CSIC; U. Autónoma de Madrid, E-28049 Madrid, Spain.}

\contribID{129}

\doi  

\maketitle

\begin{abstract}
Lipatov's high-energy effective action is a useful tool for computations in the Regge limit beyond leading order.  Recently, a regularisation/subtraction prescription has been proposed that allows to apply this formalism to calculate next-to-leading order corrections in a consistent way. 
We illustrate this procedure with the computation of the gluon Regge trajectory at two loops.
\end{abstract}

\section{The High-Energy Effective Action (HEA)}

Effective field theories provide a useful framework to treat problems involving a hierarchy of scales in quantum field theory, and are widely used in the context of QCD (e.g. chiral perturbation theory or heavy-quark effective theory). A hierarchy of scales is also present in the Regge or high-energy limit, since the centre-of-mass energy
squared $s$ is asymptotically larger than the momentum transfer $-t$ in a scattering process, hence we expect the effective theory approach to be applicable to this case.
Besides making computations simpler, a hermitian HEA incorporates unitarity. The reggeized gluon~\cite{Lipatov:1976zz} plays a key role  as the effective degree of freedom. The HEA has already been used to calculate reggeon vertices~\cite{Antonov:2004hh,Braun:2006sk}. It was introduced by Lipatov for leading order computations~\cite{Lipatov:1991nf}. 

\begin{figure}[htb]
    \label{fig:subfigures}
    \begin{center}
        \subfigure[]{%
            \label{fig:first}
            \includegraphics[width=0.3\textwidth]{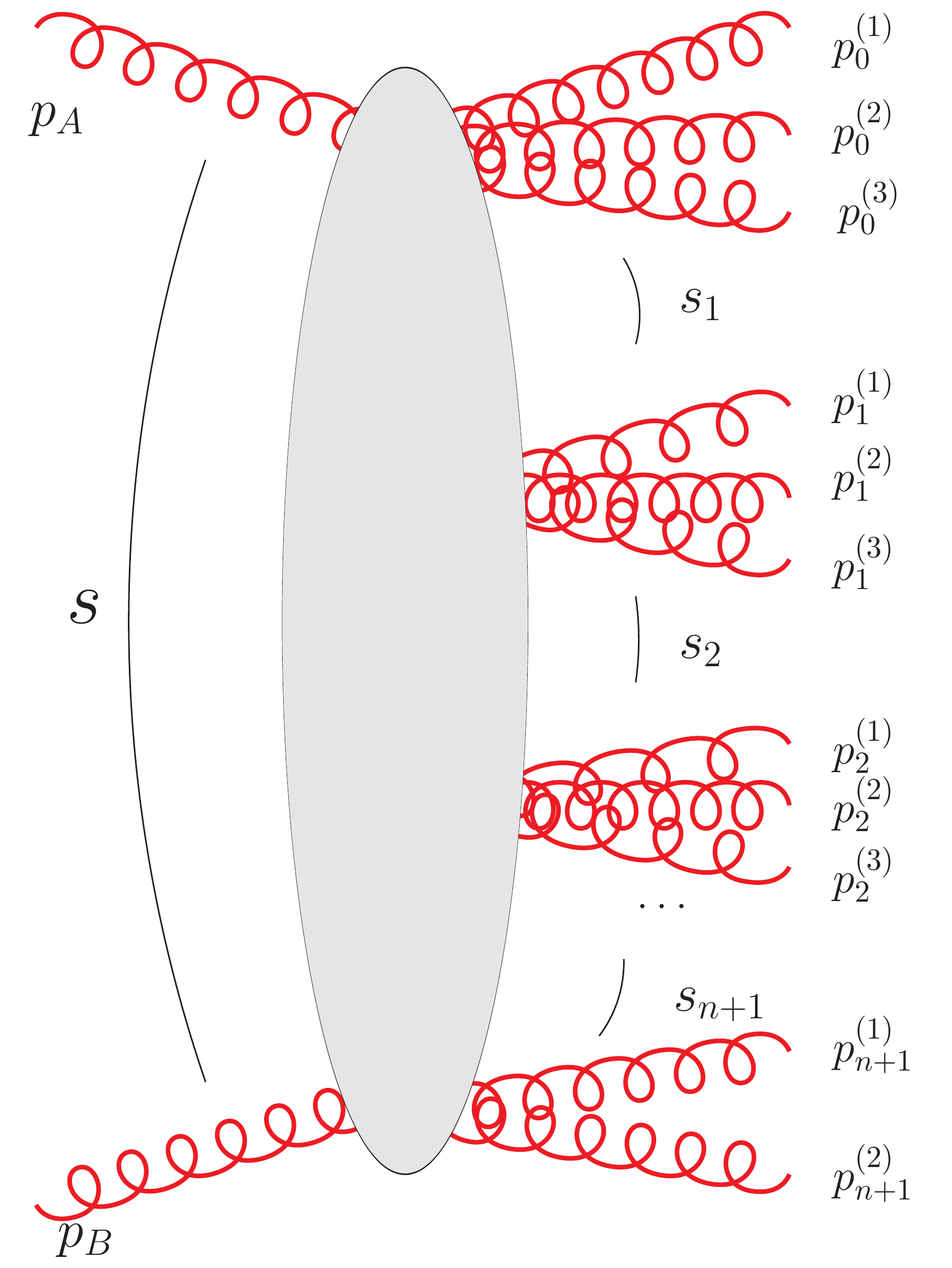}
        }%
        \subfigure[]{%
           \label{fig:second}
           \includegraphics[width=0.4\textwidth]{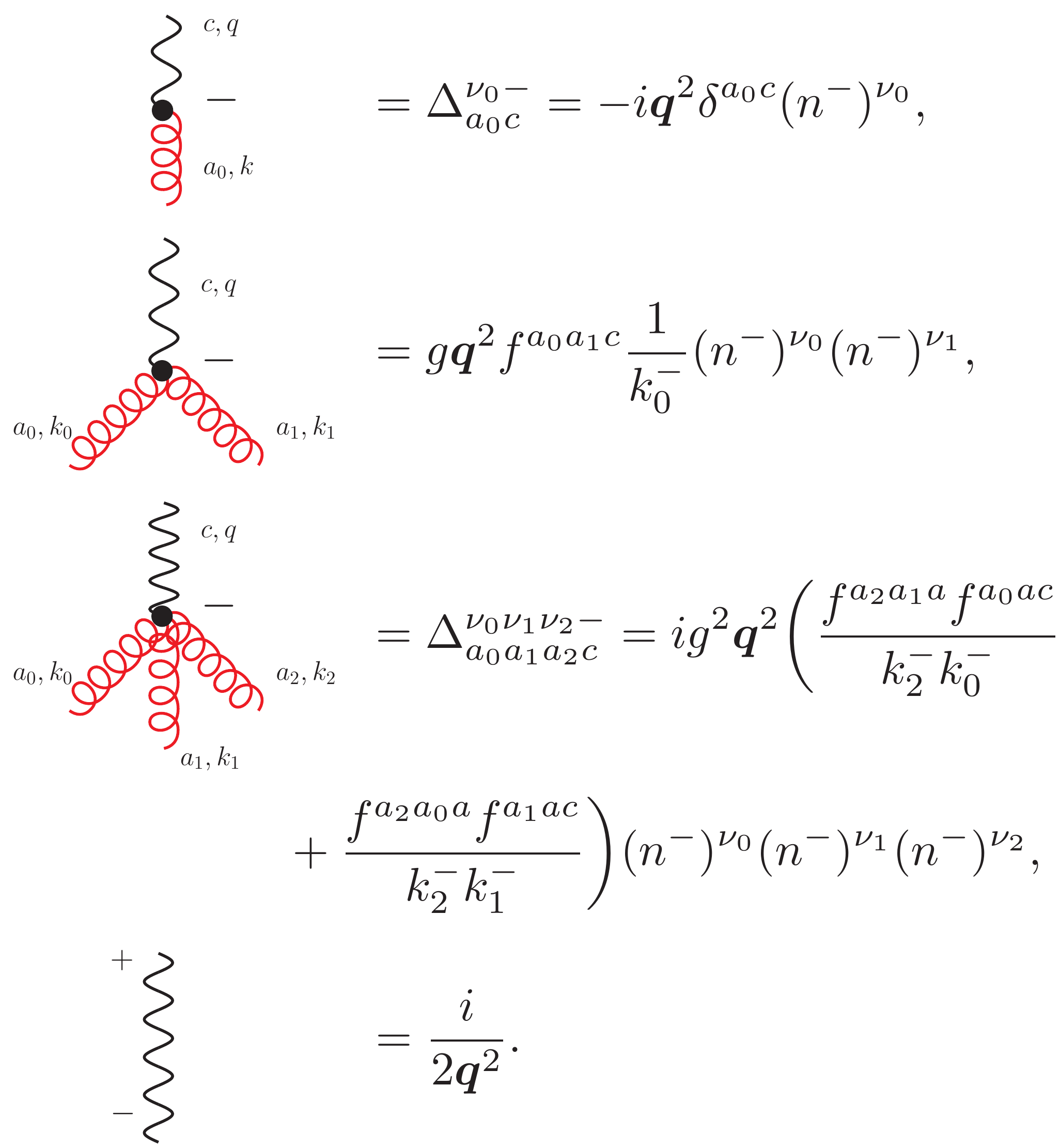}
        }\\ 
        \subfigure[]{%
            \label{fig:third}
            \includegraphics[width=0.5\textwidth]{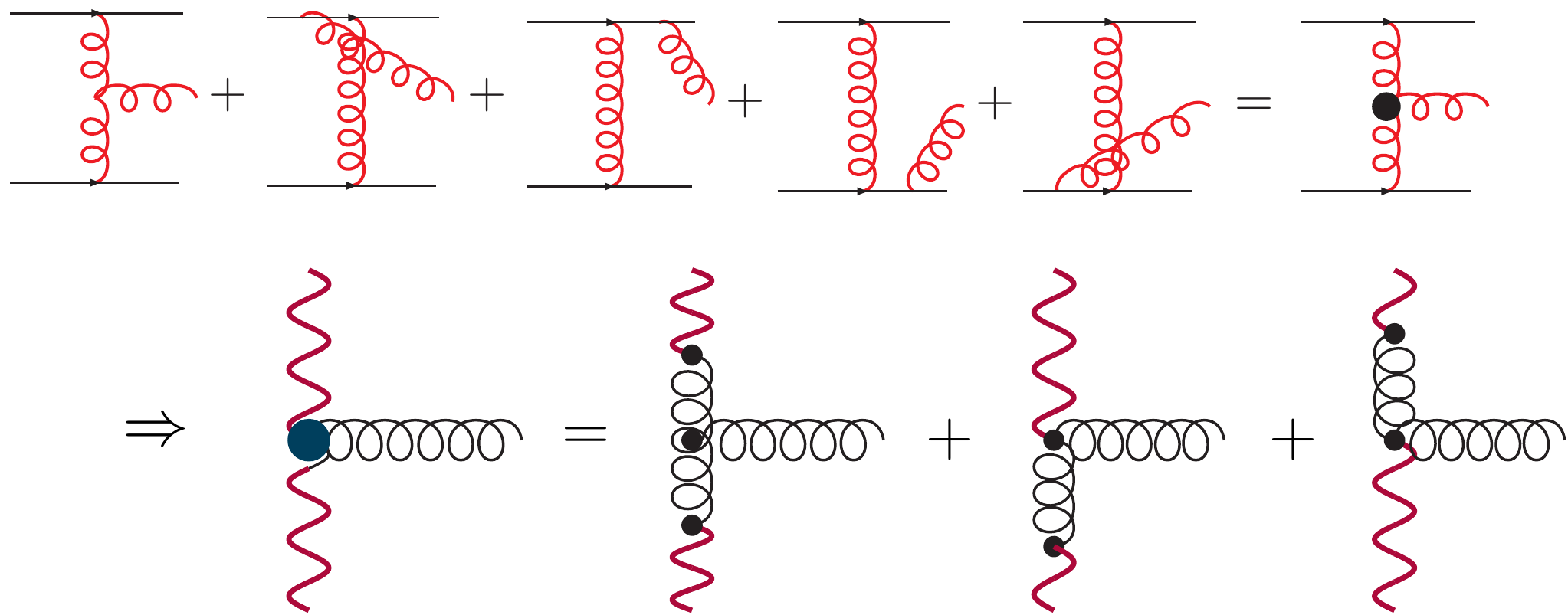}
        }%
        \subfigure[]{%
            \label{fig:fourth}
            \includegraphics[width=0.4\textwidth]{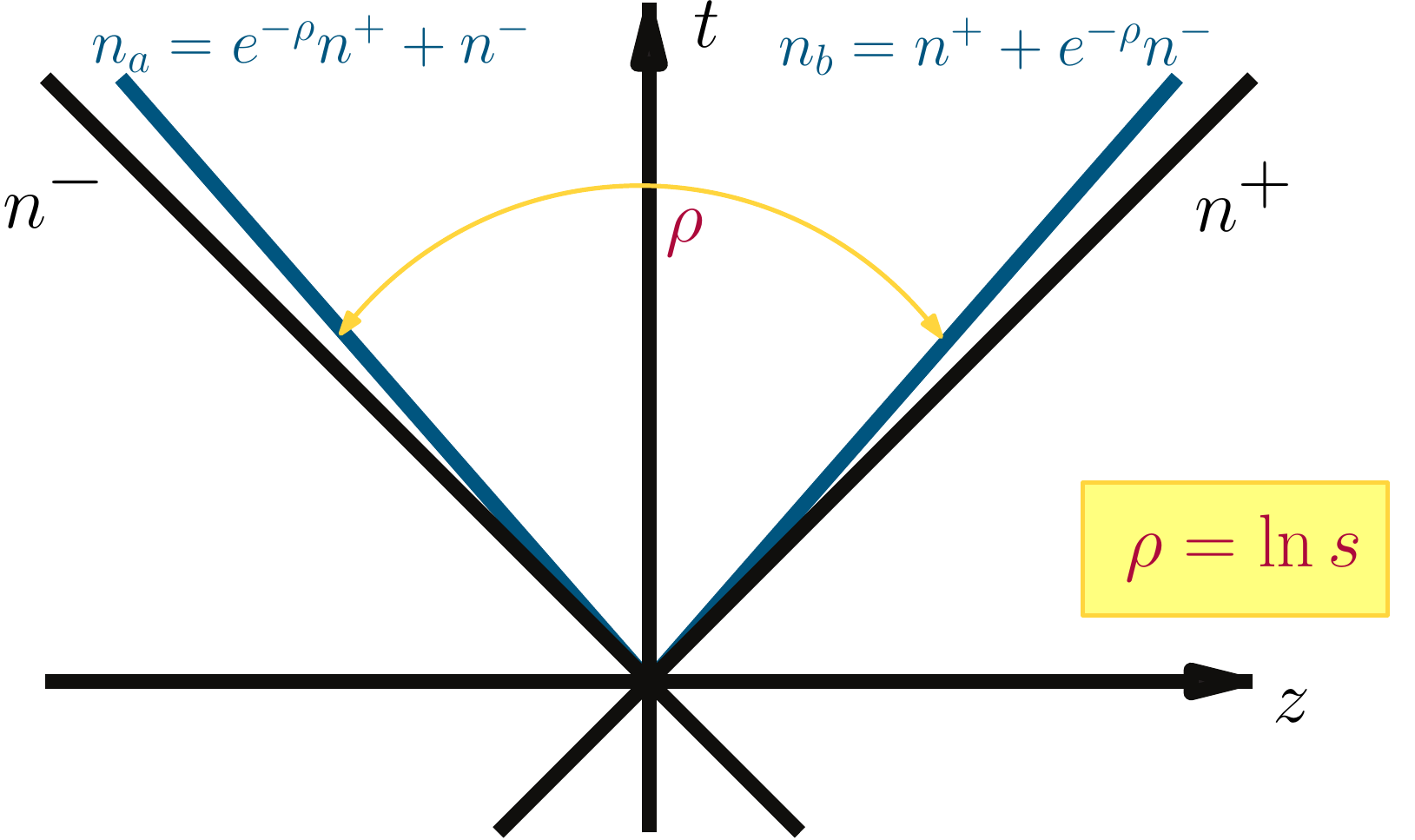}
        }%
    \end{center}
 \vspace{-0.6cm}
\caption{(a) quasi-multi-Regge kinematics; (b) Feynman rules for low-order effective vertices; (c) Reggeon-Reggeon-Gluon effective vertex; (d) Light-cone tilting regularisation.}
\label{diverse}
\end{figure}

We work with a HEA~\cite{Lipatov:1995pn} which is gauge invariant and valid beyond multi-Regge kinematics, as it includes interactions of arbitrary numbers of reggeized gluons with QCD particles. The procedure to compute 
loop calculations is explained in the following~\cite{Hentschinski:2011tz,Chachamis:2012}. Interactions take place in quasi-multi-Regge kinematics (Fig. \ref{fig:first})~\cite{Fadin:1989kf}. Emissions gather  in different clusters strongly ordered in rapidity, $y_0\gg y_1\gg\cdots\gg y_{n+1},\,y_k=\frac{1}{2}\ln\frac{k^+}{k^-}$, while all particles produced in each cluster have approximately the same rapidity. This strong ordering simplifies the polarisation tensor of $t$-channel reggeized gluons, $g_{\mu\nu}\to\frac{1}{2}(n^+)_\mu (n^-)_\nu+{\cal O}(s^{-1})$, and makes their propagators essentially transverse, $q_i^2=-\vec{q}_i^2$. Light-cone vectors are defined by $n^{+,-}=2p_{A,B}/\sqrt{s},\,k=\frac{1}{2}(k^+n^-+k^-n^+)+\vec{k}$.

Effective vertices between reggeized quarks and gluons and particles, like the one shown in Fig. \ref{fig:third}, consist of two pieces: a projection of the usual QCD vertex on QMRK, and a so-called induced contribution. This structure is given by 
\begin{equation}\label{ea}
S_{\rm eff}=S_{\rm QCD}+S_{\rm ind};\quad S_{\rm ind}=\int d^4x\, {\rm Tr}[(W_+[v(x)]-\mathscr{A}_+(x))\partial_\perp^2\mathscr{A}_-(x)]+\{+\leftrightarrow -\},
\end{equation}
where $\mathscr{A}_\pm$ are the gauge-invariant reggeon fields, which interact non-trivially with gauge invariant currents of quark and gluon fields that can be written in terms of Wilson lines $W_\pm [v]=v_\pm\frac{1}{D_\pm}\partial_\pm=v_\pm-gv_\pm\frac{1}{\partial_\pm}v_\pm+\cdots$. The projection of the reggeon polarisation tensor translates into the kinematical constraints $\partial_\pm\mathscr{A}_\mp (x)=0$ and $\sum_{i=0}^r k_i^\pm=0.$

\section{Regularisation and Subtraction}

The Feynman rules for Lipatov's HEA \cite{Antonov:2004hh} are shown in Fig. \ref{fig:second}. Poles of the form $1/k^\pm$, coming from the non-local operator $1/\partial_\pm$, are ubiquitous, and a prescription to regulate them must be taken, since they cause divergences in the longitudinal sector of loop integrals. A tilting of the light-cone by a hyperbolic angle $\rho$ was chosen in \cite{Hentschinski:2011tz}(Fig. \ref{fig:fourth}).\footnote{These poles can be considered as principal values \cite{Hentschinski:2011xg}. This does not affect the terms proportional to $\ln s$, but it is necessary for instance to recover the subleading pieces.} A technicality when using the HEA beyond tree-level is that the locality in rapidity, assumed in the derivation of \eqref{ea}, must be now enforced by hand. An alternative to the imposition of a rapidity cutoff \cite{Hentschinski:2008im}, is the subtraction of non-local contributions, mediated by reggeon exchange (see, e.g. Fig. \ref{diverse2}).

\section{Computation of NLO Gluon Regge Trajectory}

\begin{figure}[htb]
    \label{fig:subfigures}
    \begin{center}
        \subfigure[]{%
            \label{fig:first2}
            \includegraphics[width=0.6\textwidth]{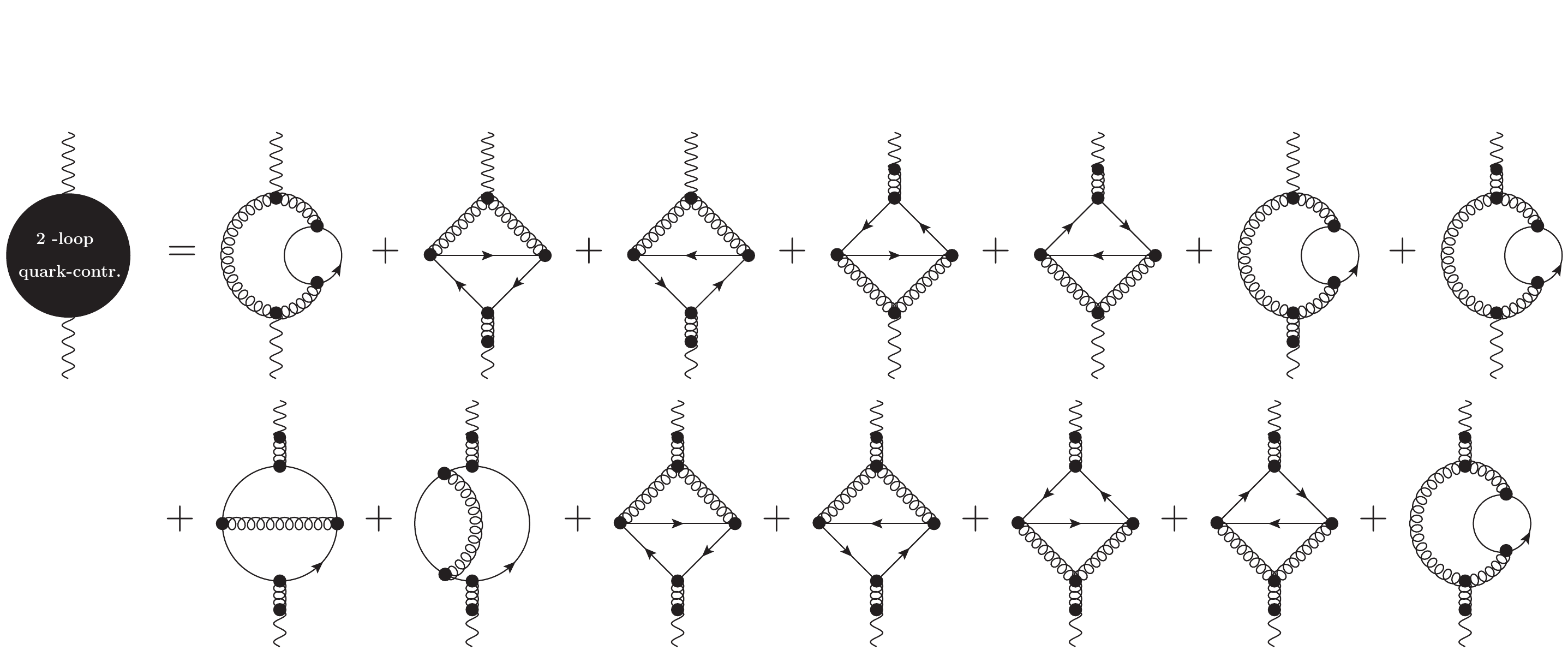}
        }%
        \subfigure[]{%
           \label{fig:second2}
           \includegraphics[width=0.3\textwidth]{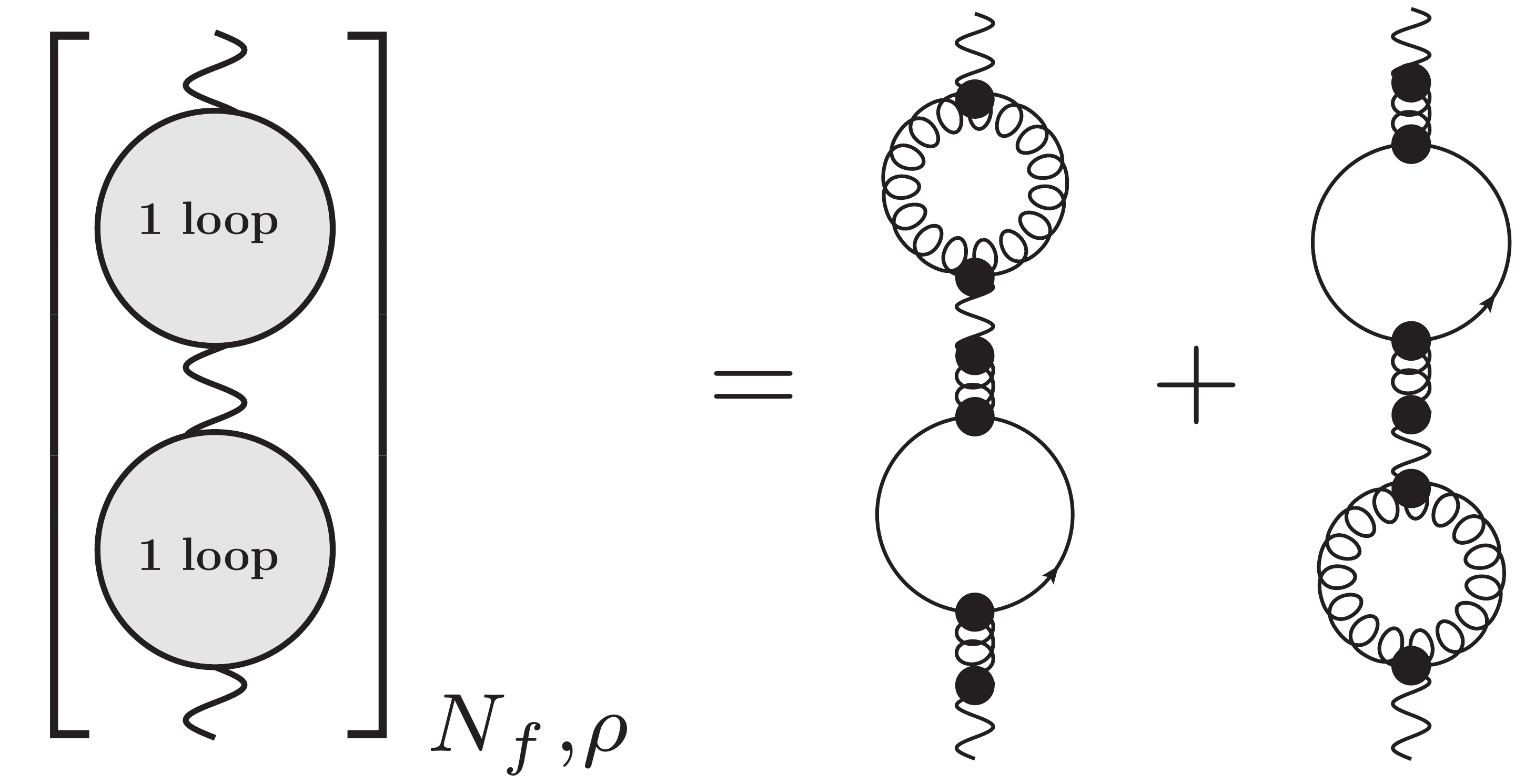}
        }
    \end{center}
 \vspace{-0.6cm}
\caption{Contributions to the quark piece of the 2-loop gluon trajectory: (a) unsubtracted gluon self-energy (only the first diagram is $\rho$-enhanced); (b) subtractions.}
\label{diverse2}
\end{figure}

This regularisation/subtraction procedure was put into work in \cite{Chachamis:2012} with the computation of the quark piece of the NLO gluon Regge trajectory, already known in QCD \cite{Fadin:1996tb} and ${\cal N}=4$ SYM \cite{Kotikov:2000pm}. The Regge trajectory is the factor $\omega(t)$ appearing in the effective propagators of reggeons, $s^{\omega(t)}$, and is a key piece in the BFKL evolution equation \cite{Lipatov:1976zz}, related to the virtual contributions. In the HEA framework, it corresponds to the diagrams in Fig. \ref{diverse2}.
\vspace{-0.6cm}

In order to compute the 2-loop gluon trajectory $\omega^{(2)}$, the following steps must be carried out \cite{Chachamis:2012}: 1) determine the high-energy limit  of the 2-loop parton-parton scattering amplitude by dropping terms suppressed when $\rho\to\infty$; 2) subtract non-local contributions to reggeised gluon self-energy; 3) divide by the tree-level result; 4) remove all terms corresponding to combinations of 1-loop trajectory and 1-loop impact factors; 5) remove a term $\frac{1}{2}\ln^2(s/s_0)[\omega^{(1)}(t)]^2$. With some modifications, the procedure is general for any other computation. For the quark piece of the two-loop trajectory, $\omega^{(2)}_q$, actually only two diagrams are $\rho$-enhanced and must be computed (one of them is the subtraction). With no rapidity cutoff, the usual techniques for computing loop integrals can be applied, and one finds exact agreement with the result in the literature
\begin{equation}
\omega^{(2)}_q\left(\frac{\vec{q}^2}{\mu^2}\right)=\bar{g}^4\left(\frac{\vec{q}^2}{\mu^2}\right)^{2\epsilon}\frac{4N_f}{\epsilon N_c}\frac{\Gamma^2(2+\epsilon)}{\Gamma(4+2\epsilon)}\left[\frac{2\Gamma^2(1+\epsilon)}{\epsilon\Gamma(1+2\epsilon)}-\frac{3\Gamma(1-2\epsilon)\Gamma(1+\epsilon)\Gamma(1+2\epsilon)}{\epsilon\Gamma^2(1-\epsilon)\Gamma(1+3\epsilon)}\right];
\end{equation}
with $\bar{g}^2=\frac{g^2N_c\Gamma(1-\epsilon)}{(4\pi)^{2+\epsilon}}$ and $d=4+2\epsilon$.\\

The gluon piece of $\omega^{(2)}$ is currently under study. More powerful technology is needed there to reduce the diagrams to master integrals and to compute the integrals themselves. It is expected however that future developments along these lines will make Lipatov's action become a useful tool for computations in the Regge limit.\\

{\footnotesize Research supported by E. Comission [LHCPhenoNet
(PITN-GA-2010-264564)] \& C. Madrid (HEPHACOS
ESP-1473).} 

{\raggedright
\begin{footnotesize}



\end{footnotesize}
}


\end{document}